\documentclass[12pt]{iopart}

\expandafter\let\csname equation*\endcsname\relax
\expandafter\let\csname endequation*\endcsname\relax
\usepackage{amsmath}
\usepackage{amssymb}
\usepackage{mathtools}
\usepackage{iopams}

\begin{document}

\title[Competition of self- and other-regarding preferences]{Competition between self- and other-regarding preferences in resolving social dilemmas}

\author{Chaoqian Wang$^{1}$ and Attila Szolnoki$^{2,*}$}
\address{$1$ Department of Computational and Data Sciences, George Mason University, Fairfax, VA 22030, USA\\
$2$ Institute of Technical Physics and Materials Science, Centre for Energy Research, P.O. Box 49, H-1525 Budapest, Hungary\\
$*$ Corresponding author}
\ead{\\CqWang814921147@outlook.com (Chaoqian Wang)\\ szolnoki.attila@ek-cer.hu (Attila Szolnoki)}

\vspace{10pt}
\begin{indented}
\item[]
\end{indented}

\begin{abstract}
Evolutionary game theory assumes that individuals maximize their benefits when choosing strategies. However, an alternative perspective proposes that individuals seek to maximize the benefits of others. To explore the relationship between these perspectives, we develop a model where self- and other-regarding preferences compete in public goods games. We find that other-regarding preferences are more effective in promoting cooperation, even when self-regarding preferences are more productive. Cooperators with different preferences can coexist in a new phase where two classic solutions invade each other, resulting in a dynamical equilibrium. As a consequence, a lower productivity of self-regarding cooperation can provide a higher cooperation level. Our results, which are also valid in a well-mixed population, may explain why other-regarding preferences could be a viable and frequently observed attitude in human society.
\end{abstract}


\section{Introduction}\label{secintro}
How can cooperation survive if defection promises a larger individual income? This is the central  question of evolutionary game theory, which seeks to understand the subtle interactions among individuals with conflicting interests~\cite{weibull_95,sigmund2010calculus}. In a simplified situation, individuals face two choices: contributing to a common pool or not. The decision determines their individual payoff and, consequently, their success. This scenario forms the core of the public goods game, where individual and collective interests are in conflict within a group~\cite{perc_jrsi13,javarone_epl16,wang2022reversed}. According to the principle of Darwinian selection, a successful strategy spreads more easily, leading defection to become the dominant strategy among selfish individuals~\cite{maynard_82}. The resulting evolutionary process would lead to the so-called ``Tragedy of the Commons''~\cite{hardin1968tragedy}, which is frequently against our real-life experiences~\cite{nowak2006evolutionary,nowak_11}.

Over the last decades, several insightful approaches have been proposed to unravel this mystery. These solutions encompass sophisticated strategies~\cite{nowak_n93,press_pnas12}, the presence of various mechanisms~\cite{nowak2006five,amaral_pre18,cardinot_njp19,chen_xj_pa08,wang_cq_amc23,liang_rh_pre22}, other incentives to support cooperation~\cite{helbing_ploscb10,brandt_pnas06,javarone_jsm24,szolnoki_epl10}, and specific conditions, such as the extensive application of network theory in structured populations~\cite{nowak1992evolutionary,santos_prl05,fu_pla07,floria_pre09,wang_z_epjb15,lieberman2005evolutionary,allen2017evolutionary}. It is almost impossible to list all of them in a short Introduction. Instead, readers seeking more comprehensive information are directed to review papers that cover these topics in detail~\cite{szabo_pr07,perc2017statistical}. The focus of the present work is on a significant, though somewhat overlooked, aspect: when individuals choose not to maximize their own payoff. Instead, the modified goal is to improve the collective income of co-players~\cite{frohlich_pc04,szabo_jtb12,pei_hy_cpb21}. 

This other-regarding preference is frequently observed not only in human society but also across the animal kingdom~\cite{burkart_pnas07,grund_srep13,platkowski_pa22,han2023novel}. The mentioned behavior was not simply studied by model calculations but also comprehensively reviewed from an economic perspective~\cite{mitteldorf_jtb00,dufwenberg_res11,wang_z_pa11,li_q_c22,zhang_yl_tcss24}. Apparently, if all participants adopt this approach, cooperators stand a better chance of survival. However, the critical question remains whether such a strategy updating preference can emerge and persist as a result of evolutionary processes. To address this, we propose a model in which both traditional self-regarding players and the introduced other-regarding ones are present. Both defectors and cooperators can exhibit these preferences, leading to a four-profile model. We systematically explore the full parameter space in this modified public goods game model, where the collective productivity of these preferences vary. 

We find that other-regarding preferences are more effective in promoting cooperation than self-regarding preferences at equally strong productivities. Importantly, in a specific parameter region where the productivities of two preferences differ, a new phase arises where cooperators with both preferences coexist. When the system is in this phase, a lower productivity of self-regarding cooperators can result in a higher general cooperation level. To check the robustness of our observations, we also study well-mixed populations, where the conclusions remain consistent. 

The organization of this paper is as follows. We first present the model in Sec.~\ref{sec_model} and then proceed with our observations and their explanations in Sec.~\ref{sec_results}. Section~\ref{sec_concl} contains our conclusions and a discussion of their implications. Last, we end with \ref{A} where the details of calculations in a well-mixed population are provided.

\section{Model}\label{sec_model}
In the extended model, players are categorized based on two unconditional strategies, cooperation ($C$) and defection ($D$), used in the classic model, as well as two preferences in updating strategies: self-regarding ($0$) and other-regarding ($1$). In this way, there are four profiles: cooperation and self-regarding preference ($C_0$), cooperation and other-regarding preference ($C_1$), defection and self-regarding preference ($D_0$), and finally defection and other-regarding preference ($D_1$). Technically each player's profile is denoted by a vector, $\mathbf{s}_i=(s^{[C_0]}_i,s^{[C_1]}_i,s^{[D_0]}_i,s^{[D_1]}_i)$. The element corresponding to the player's profile is set to 1, while the remaining three components are 0. For example, if an agent adopts cooperation and self-regarding preference, its profile is represented as $\mathbf{s}_i=(1,0,0,0)$, whereas the profile of a cooperator with other-regarding preference is represented as $\mathbf{s}_i=(0,1,0,0)$, and so on.

In the spatial public goods game framework~\cite{szolnoki_pre09c,flores_jtb21,quan_j_csf21,wang_jw_pla22,yu_fy_csf22,sun_xp_pla23}, we consider a square lattice of size $L\times L$ and population $N=L^2$, where each agent $i$ occupies a node and forms a group $\Omega_i$ with size $G=5$, consisting of itself and its four nearest neighbors. Consequently, agent $i$ is also a member of the groups $\Omega_j$ of its neighbors $j\in\Omega_i\setminus \{i\}$. In the public goods game of the group centered on $j$, agent $i$ contributes a cost $c>0$ if cooperating or contributes nothing if defecting. The contributions from all group members $k\in\Omega_j$ are enhanced by a synergy or productivity factor $r>1$ and evenly distributed among all $G$ members. Furthermore, the payoff $\pi_i^{(0)}$ of agent $i$ is the average over the games played in the $G$ groups, given by
\begin{equation}\label{eq_pi0}
\pi_i^{(0)}=\frac{1}{G}\sum_{j\in \Omega_i}\left(\frac{1}{G}\sum_{k\in \Omega_j}(r_0 s^{[C_0]}_k c+r_1 s^{[C_1]}_k c)-(s^{[C_0]}_i+s^{[C_1]}_i)c\right).
\end{equation}
To distinguish between agents with self- and other-regarding preferences, we introduced a technical modification in Eq.~(\ref{eq_pi0}), where the diverse productivity of cooperation with self- and other-regarding learning are characterized by $r_0$ and $r_1$ (both $r_0>1$ and $r_1>1$), respectively. 

The average payoff of agent $i$'s nearest neighbors is denoted by $\pi_i^{(1)}$ and is calculated as
\begin{equation}\label{eq_orpayoff}
\pi_i^{(1)}=\frac{1}{G-1}\sum_{j\in \Omega_i\setminus \{i\}}\pi_j^{(0)}.
\end{equation}

If agent $i$ adopts other-regarding preference, it rescales the original payoff by weighing its own payoff and the payoff of neighbors. The payoff of $i$'s neighbors is weighted by $0\leq u\leq 1$, and the payoff of agent $i$ is weighted by $1-u$, where $u$ is the other-regarding rate. Accordingly, the rescaled payoff is denoted by $\tilde{\pi}_i^{(1)}$, which is given by
\begin{equation}\label{eq_tildepi1}
\tilde{\pi}_i^{(1)}=(1-u)\pi_i^{(0)}+u\pi_i^{(1)}.
\end{equation}

An agent $i$ pursues a higher $\tilde{\pi}_i^{(1)}$ when engaging in other-regarding learning, and prefers a higher $\pi_i^{(0)}$ when engaging in self-regarding learning. This is captured by the following updating process~\cite{szabo1998evolutionary}. At each elementary Monte Carlo step, a random agent $i$ and a random one of its neighbors $i' \in \Omega_i \setminus \{i\}$ are selected, and their payoffs are calculated. If agent $i$ has other-regarding preference (i.e., $\mathbf{s}_i=(0,1,0,0)$ or $\mathbf{s}_i=(0,0,0,1)$), then $i$ adopts the profile of $i'$ with probability
\begin{equation}\label{eq_fermi1}
\tilde{P}_1(\mathbf{s}_i \gets \mathbf{s}_{i'}) = \frac{1}{1+\exp{[-(\tilde{\pi}_{i'}^{(1)}-\tilde{\pi}_{i}^{(1)})/\kappa]}}.
\end{equation}
If agent $i$ has self-regarding preference (i.e., $\mathbf{s}_i=(1,0,0,0)$ or $\mathbf{s}_i=(0,0,1,0)$), then $i$ adopts the profile of $i'$ with probability
\begin{equation}\label{eq_fermi0}
P_0(\mathbf{s}_i \gets \mathbf{s}_{i'}) = \frac{1}{1+\exp{[-(\pi_{i'}^{(0)}-\pi_{i}^{(0)})/\kappa]}}.
\end{equation}

In Eqs.~(\ref{eq_fermi1}) and (\ref{eq_fermi0}), a higher $\tilde{\pi}_{i'}^{(1)}$ or $\pi_{i'}^{(0)}$ makes it more likely for agent $i$ to imitate the profile of $i'$. To gain results comparable with previous studies~\cite{szolnoki_pre09c,flores_pre23,zhang_h_csf23,duan_yx_csf23,quan_j_c23}, we set $\kappa=0.1$, the noise parameter in both probability functions.
Importantly, the motivation of agents is complex: the update happens to the entire profile vector $\mathbf{s}_i \gets \mathbf{s}_{i'}$. If agent $i$ has other-regarding preference, it compares the rescaled neighboring payoff between the reference neighbor and itself, in an attempt to learn the profile that brings others higher payoff. This is independent of the reference profile: if ``self-regarding preference'' is found to bring others a higher payoff, then agent $i$ switches to ``self-regarding preference'' in pursuit of its current purpose of other-regarding preference. Similar reasoning holds if agent $i$ adopts self-regarding preference: when the reference neighbor $i'$ has a high personal payoff with other-regarding preference, agent $i$ adopts both the strategy and other-regarding preference from $i'$ in order to achieve the purpose of a higher personal payoff.

The model can reduce to the classic spatial public goods game when $r_0=r_1\equiv r$ in Eq.~(\ref{eq_pi0}) and $u=0$ in Eq.~(\ref{eq_tildepi1}). If $r_0=r_1$, there is no difference between the productivities of self- and other-regarding cooperation. If $u=0$, we have $\tilde{\pi}_i^{(1)}=\pi_i^{(0)}$ and $\tilde{P}_1(\mathbf{s}_i\gets \mathbf{s}_{i'})=P_0(\mathbf{s}_i\gets \mathbf{s}_{i'})$. The other-regarding preference is not truly other-regarding and becomes indistinguishable from self-regarding preference. If $u=1$, we have $\tilde{\pi}_i^{(1)}=\pi_i^{(1)}$. The modified preference is completely other-regarding, and these agents only pursue the higher payoff of neighbors. Overall, investigating $0<u<1$ allows for an examination of the gradual change of other-regarding agents from self-regarding to complete other-regarding preferences, and studying $r_0\neq r_1$ can provide additional insights by directly adjusting the relative advantages of each.

To keep the results comparable with previous studies, some parameters, including group size $G=5$, cost $c=1$, and selection noise $\kappa=0.1$, are fixed in the simulations. The typical population size is $L\times L=300\times 300$, which allows us to avoid finite-size effect. The evolution starts from a state where the four profiles are distributed randomly. The total running time is at least $10^5$ full Monte Carlo steps where the stationary quantities are averaged in the last $2\times 10^4$ full Monte Carlo steps. The measured quantities are as follows: the fraction of self-regarding cooperation $\rho_{C_0}=\sum_{i=1}^{N} s_i^{[C_0]}/N$, the fraction of other-regarding cooperation $\rho_{C_1}=\sum_{i=1}^{N} s_i^{[C_1]}/N$, the fraction of self-regarding defection $\rho_{D_0}=\sum_{i=1}^{N} s_i^{[D_0]}/N$, the fraction of other-regarding defection $\rho_{D_1}=\sum_{i=1}^{N} s_i^{[D_1]}/N$, and the fraction of cooperation $\rho_C=\rho_{C_0}+\rho_{C_1}=\sum_{i=1}^{N} (s_i^{[C_0]}+s_i^{[C_1]})/N$. The presented stationary quantities are robust by changing system size and simulation time. To close the description of our model definition, we note that it is possible to use an alternative model setup where pure strategies and other player-specific tags can have different time scales to change~\cite{roca_prl06,szolnoki_epjb09}. However, in our present model, such an extension would cause unnecessary complexity; hence, we keep the original setup focused on the proper competition of self- and other-regarding preferences.

\section{Results and discussion}\label{sec_results}

\subsection{Equally strong preferences}
\begin{figure}
	\centering
		\includegraphics[width=.6\textwidth]{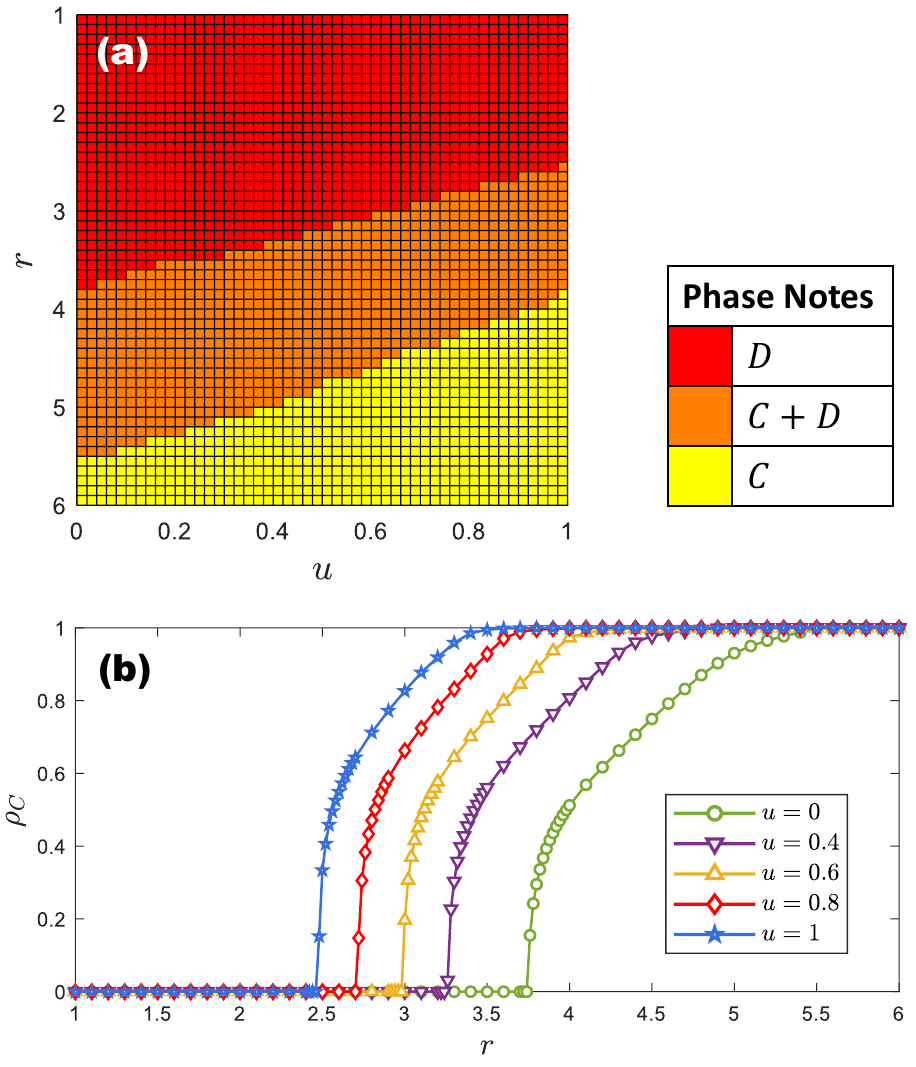}
	\caption{Panel~(a): phase diagram on the parameter plane of productivity factor $r_0=r_1\equiv r$ and other-regarding rate $u$. The full defection and cooperation phases are separated by a mixed phase, conceptually similar to the traditional PGG model. Panel~(b): the fraction of cooperation $\rho_C$ as a function of productivity $r_0=r_1\equiv r$ at different other-regarding rate $u$. An increase in other-regarding rate $u$ supports general cooperation. } 
	\label{fig_ru}
\end{figure}

As a natural entry point, we study a special case where the productivities of self- and other-regarding cooperation are equal, $r_0=r_1\equiv r$. According to Eqs.~(\ref{eq_pi0})--(\ref{eq_tildepi1}), $\pi_i^{(0)}\neq \tilde{\pi}_i^{(1)}$ holds when $u>0$. Therefore, the learning probabilities $\tilde{P}_1(\mathbf{s}_i \gets \mathbf{s}_{i'})$ (other-regarding) and $P_0(\mathbf{s}_i \gets \mathbf{s}_{i'})$ (self-regarding), determined by Eqs.~(\ref{eq_fermi1}) and (\ref{eq_fermi0}), are distinct. However, we focus on the effect of other-regarding rate $u$ on the general cooperation level ($\rho_C$). To this end, we aggregate the fractions of cooperation, as well as defection, across both preferences. The resultant phase diagram on the parameter plane of productivity $r$ and other-regarding rate $u$ is shown in Fig.~\ref{fig_ru}(a). We can distinguish three phases here. As expected, defectors dominate at low $r$ values, regardless of $u$. At high $r$ values, cooperators prevail. Located between these solutions is a mixed $C+D$ phase where both strategies coexist. The critical $r$ value, at which cooperation becomes dominant, decreases as $u$ increases, reinforcing the expectation that other-regarding preferences support cooperation. Cross-sections obtained at different $u$ values are shown in Fig.~\ref{fig_ru}(b). These curves confirm that an increase in other-regarding rate $u$ helps cooperation even in the mixed phase at a constant $r$, which supports a previous conclusion that ``standing in others' shoes promotes cooperation''~\cite{han2023novel}. Finally, at $u=0$, we have $\pi_i^{(0)}= \tilde{\pi}_i^{(1)}$ according to Eqs.~(\ref{eq_pi0})--(\ref{eq_tildepi1}), which makes self- and other-regarding preferences indistinguishable, reducing the results to those obtained in the traditional spatial public goods game~\cite{szolnoki_pre09c}.

\begin{figure}
	\centering
		\includegraphics[width=.7\textwidth]{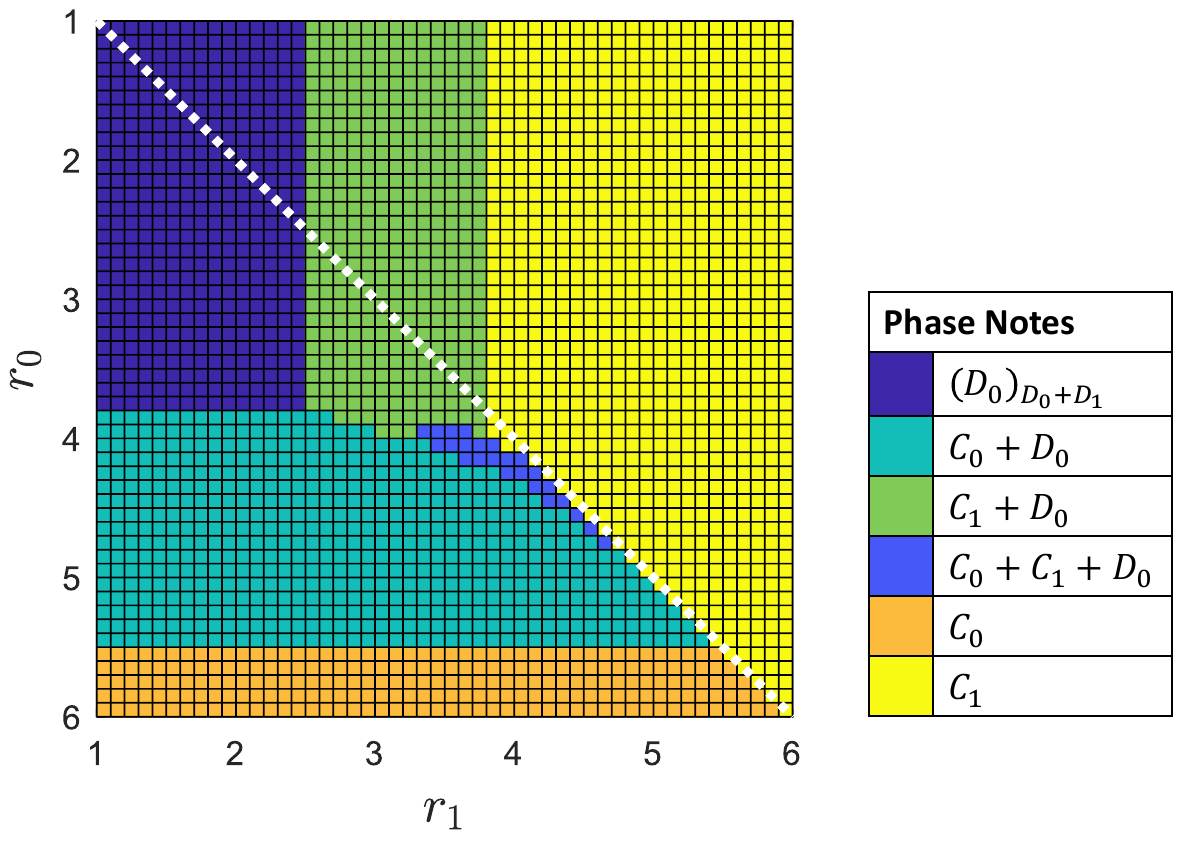}
	\caption{Phase diagram on the parameter plane of self-regarding productivity ($r_0$) and other-regarding productivity ($r_1$). The other-regarding rate is $u=1$. The white dashed line marks $r_0=r_1$. When $r_0$ and $r_1$ are largely different, we get back to the pattern in simplified two-profile models. Interestingly, other-regarding cooperation can conquer self-regarding cooperation even if $r_0 > r_1$. Furthermore, when $r_0$ slightly exceeds $r_1$, a new phase emerges composed of $C_0$, $C_1$, and $D_0$. The nature of this solution is discussed in the main text.} 
	\label{fig_phase2D_u1}
\end{figure}

\subsection{Unequally productive preferences}
A question of interest is whether other-regarding preferences can emerge in the presence of self-regrading preferences and what solutions arise when both preferences compete. To provide a general answer, we remove the artificial constraint of $r_0 = r_1$ and allow for $r_0 \neq r_1$, the productivities vary among cooperators with different preferences. First, we use $u=1$, where the potential impact of other-regarding preference is maximal. In other words, players with profiles of $(0,1,0,0)$ or $(0,0,0,1)$ completely ignore their individual income and are solely influenced by the average payoff of their neighbors.
The resulting phase diagram is shown in Fig.~\ref{fig_phase2D_u1}, where dominant solutions are marked on the $r_0$-$r_1$ plane. These solutions are color-coded and explained in ``Phase Notes.'' 

For largely different values of $r_0$ and $r_1$, the results resemble the patterns obtained in two-profile models. When the productivities for both preferences are low, the system evolves into a full defection state, marked as $(D_0)_{D_0+D_1}$. Here, the microscopic dynamics becomes a neutral drift when the last cooperator dies out because $\pi_i^{(0)}= \tilde{\pi}_i^{(1)}=0$ for all players. However, this voter-model-like coarsening is not fully symmetric because a higher starting portion for the $D_0$ profile ensures a higher fixation probability of $D_0$. This is why we denoted this phase as ``$D_0$'' with $D_0+D_1$ subscript. As $r_0$ or $r_1$ increases, the system evolves into the $C_0+D_0$ or $C_1+D_0$ phase, where self- or other-regarding cooperation coexists with self-regarding defection by spatial reciprocity. As $r_0$ or $r_1$ becomes sufficiently large, the system enters the full $C_0$ or full $C_1$ phase. It is noteworthy that the emergence of other-regarding cooperation $C_1$ requires a smaller productivity $r_1$ compared to the necessary $r_0$ for the emergence of self-regarding cooperation $C_0$. The reason is straightforward: other-regarding cooperation cannot easily transform into self-regarding defection, as the latter does not benefit neighbors.

We also note that the ``other-regarding defection'' profile cannot survive in evolution according to the full phase diagram shown in Fig.~\ref{fig_phase2D_u1}. Importantly, individuals have no preconceived moral cognition about behaviors in evolutionary dynamics. The ``defection'' and ``other-regarding'' refer to the process of playing games and updating profiles, thus belonging to independent dimensions and do not conflict. Yet, evolution does not favor this profile. This is because that the other-regarding preference aims to maximize the payoff of neighbors, which defection cannot achieve. When faced with $C_0$ or $C_1$, other-regarding defection $D_1$ easily transforms into $C_0$ or $C_1$, while the reverse process is difficult. When faced with $D_0$, the transformations follow a neutral drift, but before extinction, $C_0$ or $C_1$ players can beat most $D_1$ neighbors, leading to a larger initial fraction of $D_0$ and hence a higher fixation probability for $D_0$.

The phase diagram of Fig.~\ref{fig_phase2D_u1} also shows that the $C_1+D_0$ phase (light green) crosses the $r_0=r_1$ diagonal marked by a white dashed line. It means that other-regarding preferences can beat self-regarding ones even if the latter have higher productivity. This phenomenon is quite impressive and answers our original question: other-regarding preference can emerge even in the presence of self-regarding players because it better stabilizes the coexistence with defection especially in the low-productivity interval.

\begin{figure}
	\centering
		\includegraphics[width=.95\textwidth]{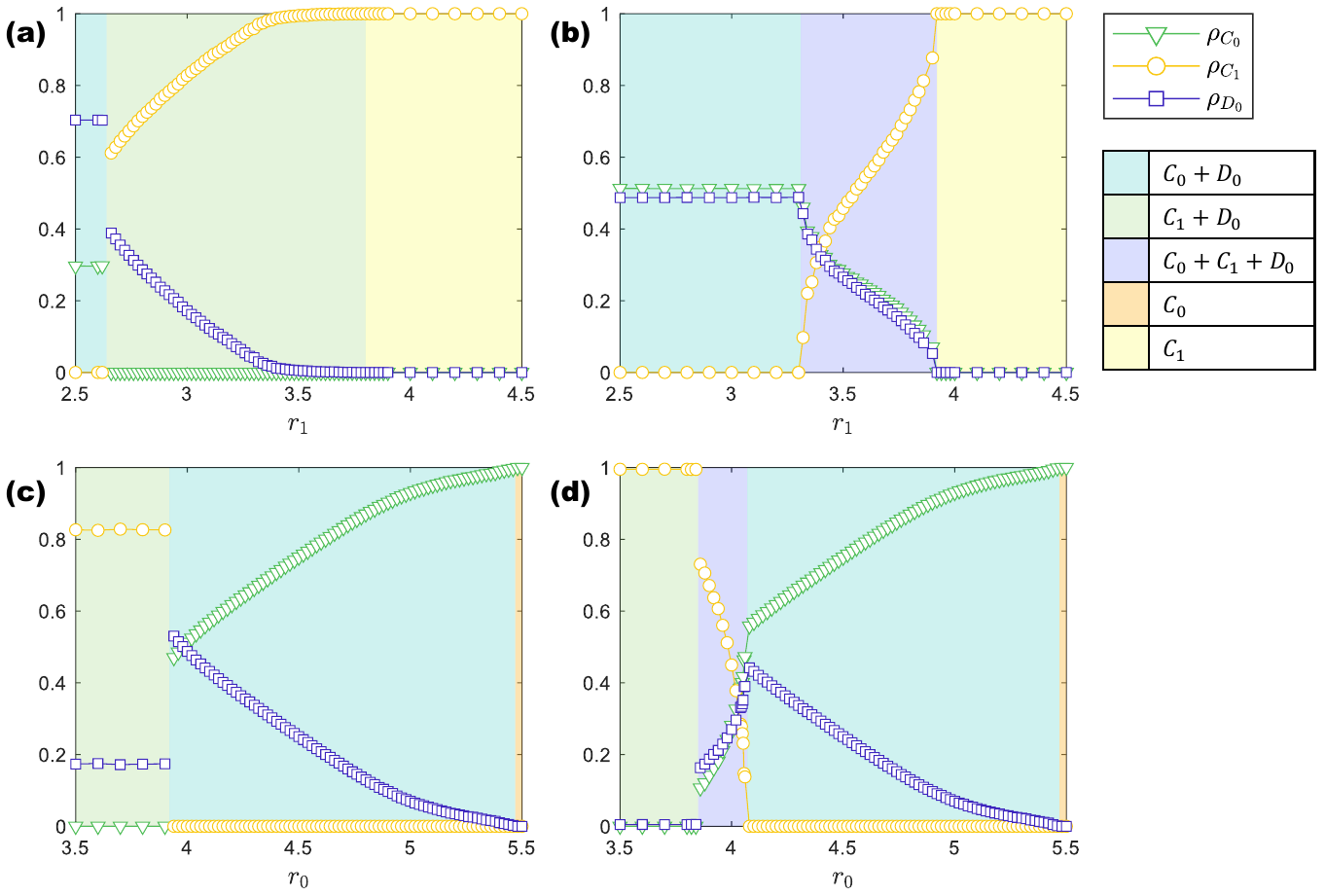}
	\caption{Horizontal and vertical cross-sections of the phase diagram of Fig.~\ref{fig_phase2D_u1} showing the stationary fractions of profiles. Panel~(a): fractions as a function of $r_1$ at $r_0=3.8$. There is a discontinuous phase transition between $C_0+D_0$, $C_1+D_0$ followed by a continuous phase transition to the $C_1$ phase. Panel~(b): fractions as a function of $r_1$ at $r_0=4$. The $C_0+D_0$ phase is replaced by $C_0+C_1+D_0$ and followed by the full $C_1$ phase, and the phase transitions are continuous. Panel~(c): fractions as a function of $r_0$ at $r_1=3$. The transition between the $C_1+D_0$ and $C_0+D_0$ phases is discontinuous. Panel~(d): fractions as a function of $r_0$ at $r_1=3.5$. The fraction of defection gradually increases through the $C_0+C_1+D_0$ phase as $r_0$ increases.} 
	\label{fig_phase1D_u1}
\end{figure}

The phase diagram reveals a new phase near the $r_0=r_1$ diagonal, where $r_0$ slightly exceeds $r_1$, and $C_0$, $C_1$, $D_0$ coexist. To get a deeper look at the phase transitions, we present some cross-sections of the phase diagram in Fig.~\ref{fig_phase1D_u1}. Here, the stationary $\rho_{C_0}$, $\rho_{C_1}$, and $\rho_{D_0}$ are shown as functions of $r_1$ or $r_0$. The top panels show two representative horizontal cross-sections. When self-regarding productivity $r_0$ is high enough to allow the emergence of self-regarding cooperation, but not sufficient to exclude defection, two scenarios can be observed. Panel~(a) shows the case when $r_0$ is relatively small ($r_0=3.8$). This productivity level allows self-regarding cooperation to coexist with defection. If we increase other-regarding productivity $r_1$, then $C_1$ players can replace $C_0$ players via a discontinuous phase transition. Notably, the fraction of $C_1$ players is significantly higher than the fraction of $C_0$ players on the other side of the transition point, which is of great importance as we will explain it later. As we increase $r_1$ in the $C_1+D_0$ phase, the system behavior is similar to the traditional two-profile model. Accordingly, the fractions of defection decrease gradually for larger $r_1$ and the system terminates in the $C_1$ phase via a continuous phase transition. Panel~(b) illustrates an alternative scenario obtained at $r_0=4$, where the self-regarding productivity ensures a relatively high cooperation level in the $C_0+D_0$ phase. Here, the increase in other-regarding productivity $r_1$ does not lead to a sudden switch to the $C_1+D_0$ solution. Instead, the larger $r_1$ supports $C_1$ players to form a solution with the other two profiles. Further enlarging $r_1$ could also be efficient in supporting $C_1$, and the system terminates in the full $C_1$ phase via a continuous phase transition.

The bottom panels of Fig.~\ref{fig_phase1D_u1} present vertical cross-sections of the phase diagram. We can see a seemingly counter-intuitive phenomenon: a higher synergy does not necessarily result in a higher cooperation level. For example, in Fig.~\ref{fig_phase1D_u1}(c), obtained at $r_1=3$, the fraction of defectors suddenly jumps as $r_0$ exceeds a critical value, and the system enters the $C_0+D_0$ phase. Naturally, a further increase in $r_0$ within this two-profile phase results in a decay of $\rho_{D_0}$, and the system finally reaches the full $C_0$ phase via a continuous phase transition. Similarly, at $r_1=3.5$ [Fig.~\ref{fig_phase1D_u1}(d)], as $r_0$ increases, the system exits the $C_1$ phase and enters the $C_0+C_1+D_0$ phase, where the defection level gradually increases with $r_0$. Further increasing $r_0$ leads to the $C_0+D_0$ phase, where we see similar system behavior discussed in panel~(c). To explain, high self-regarding productivity prevents the reproduction of other-regarding cooperation with low productivity: self-regarding cooperation with high productivity bring higher payoffs to their neighbors, such that other-regarding cooperation transform into them. However, other-regarding cooperation can better suppress defection: with low other-regarding productivity, cooperation can flourish more than with high self-regarding productivity. If self-regarding cooperation has a high productivity, then it prevents other-regarding cooperation from utilizing this ability, thus reducing the fraction of cooperation in the system. Therefore, to better cooperation, self-regarding cooperation should sometimes keep their productivity low to make way for other-regarding cooperation.

\begin{figure}
	\centering
		\includegraphics[width=\textwidth]{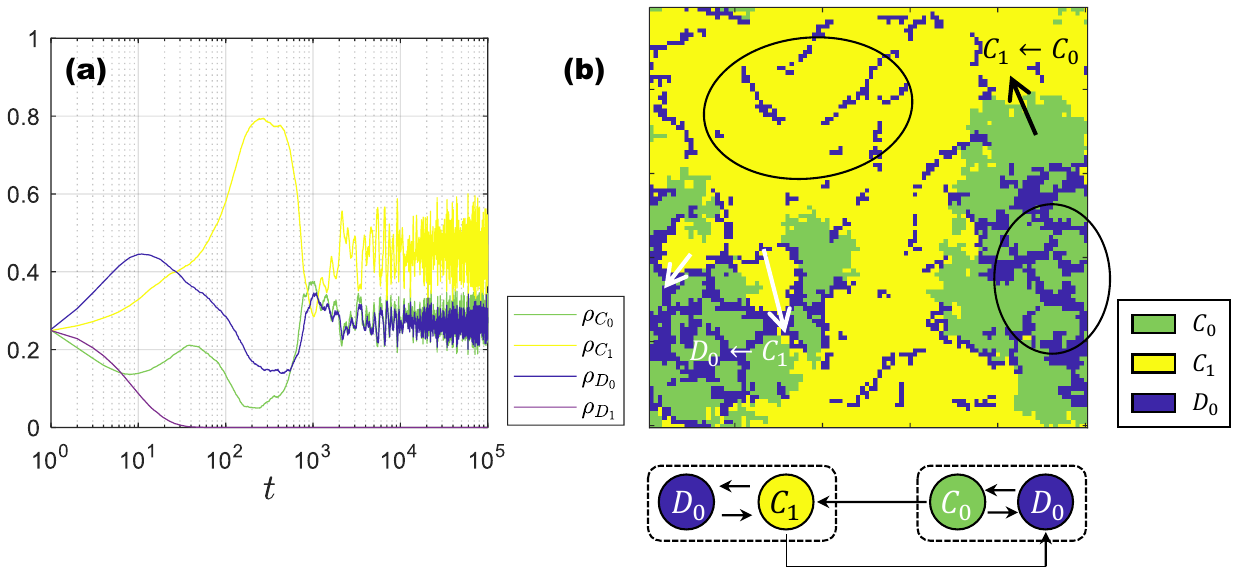}
	\caption{Two solutions forming a new one. (a) When $D_1$ dies out, the remaining three profiles form a stationary solution, as their time evolution suggests. (b) A typical snapshot of the dynamic equilibrium in the $C_0+C_1+D_0$ phase on a $100\times 100$ square lattice. Both $C_0+D_0$ and $C_1+D_0$ solutions are stationary alone as the ellipses mark. However, there are permanent and mutual invasions between these phases. The fraction of $D_0$ players differ in these solutions. $C_1$ can better suppress $D_0$ even with low productivity, thus squeezing in from large defection cracks in $C_0+D_0$ regions, transforming $C_0+D_0$ into $C_1+D_0$, shown by white arrows. $C_0$, due to high productivity, invade $C_1$, transforming $C_1+D_0$ into $C_0+D_0$, shown by a black arrow on the top. The mini diagram of these processes is on the bottom. The arrows represent the direction of invasions. Parameters: $r_0=4.0$, $r_1=3.5$, $u=1$.} 
	\label{fig_t}
\end{figure}

Next, we discuss the dynamics in the previously mentioned three-profile phase. By taking a representative combination of parameters, $r_0=4.0$, $r_1=3.5$, we present the time evolution of different profiles in Fig.~\ref{fig_t}(a). It indicates that $C_0$, $C_1$, and $D_0$ players form a stable solution after $D_1$ players die out. To stress the difference between the initial evolutionary and the final stationary state, we use semi-log plot in this panel. Panel~(b) captures a representative snapshot in the pattern formation in the stationary state. This is taken at a relatively small system size, $L \times L = 100 \times 100$, but our goal is to present the critical elements of the invasion process in detail. We use a color-coded presentation for different profiles as shown on the right-hand side. The first comment is that each of the $C_0+D_0$ and $C_1+D_0$ phases, marked by ellipses, would be a stable solution in the absence of the other at these parameter values. They are based on the network reciprocity mechanism observed by Nowak and May~\cite{nowak1992evolutionary}. It is crucial, however, that the fractions of defection differ significantly in the domains controlled by self- and other-regarding cooperation. Since other-regarding cooperation can better suppress defection even with low productivity $r_1=3.5$, only tiny ``cracks'' of defection can survive in the $C_1+D_0$ regions. Instead, in the $C_0+D_0$ regions with relatively high productivity, $r_0=4.0$, of self-regarding cooperation, defectors can exploit $C_0$ players, opening larger ``cracks.'' Other-regarding cooperation can enter these larger defection cracks and then keep smaller cracks that they can, which is the way $C_1+D_0$ regions invade $C_0+D_0$. This process is shown by white arrows in Fig.~\ref{fig_t}(b). We may say that the $D_0$ profile plays as a ``Trojan Horse'' when $C_1+D_0$ invades the area of $C_0+D_0$. Conversely, the way how $C_0+D_0$ regions invade $C_1+D_0$ is more straightforward: self-regarding cooperation $C_0$, with its higher productivity $r_0$, is able to directly beat other-regarding $C_1$ whose productivity $r_1$ is lower. This process is shown by a black arrow in Fig.~\ref{fig_t}(b). After $C_0$ agents invade $C_1$, they are not able to suppress the small cracks of defection in the previous $C_1+D_0$ regions. The uncontrolled defection expand and finally reach a balance with self-regarding cooperation, where the cracks of defection are large enough and, in turn, open up an opportunity for the invasion of $C_1$. The process is a continuous loop, forming a ``dynamic equilibrium,'' as shown in the mini diagram below Fig.~\ref{fig_t}(b).

\begin{figure}
    \centering
    \includegraphics[width=.6\textwidth]{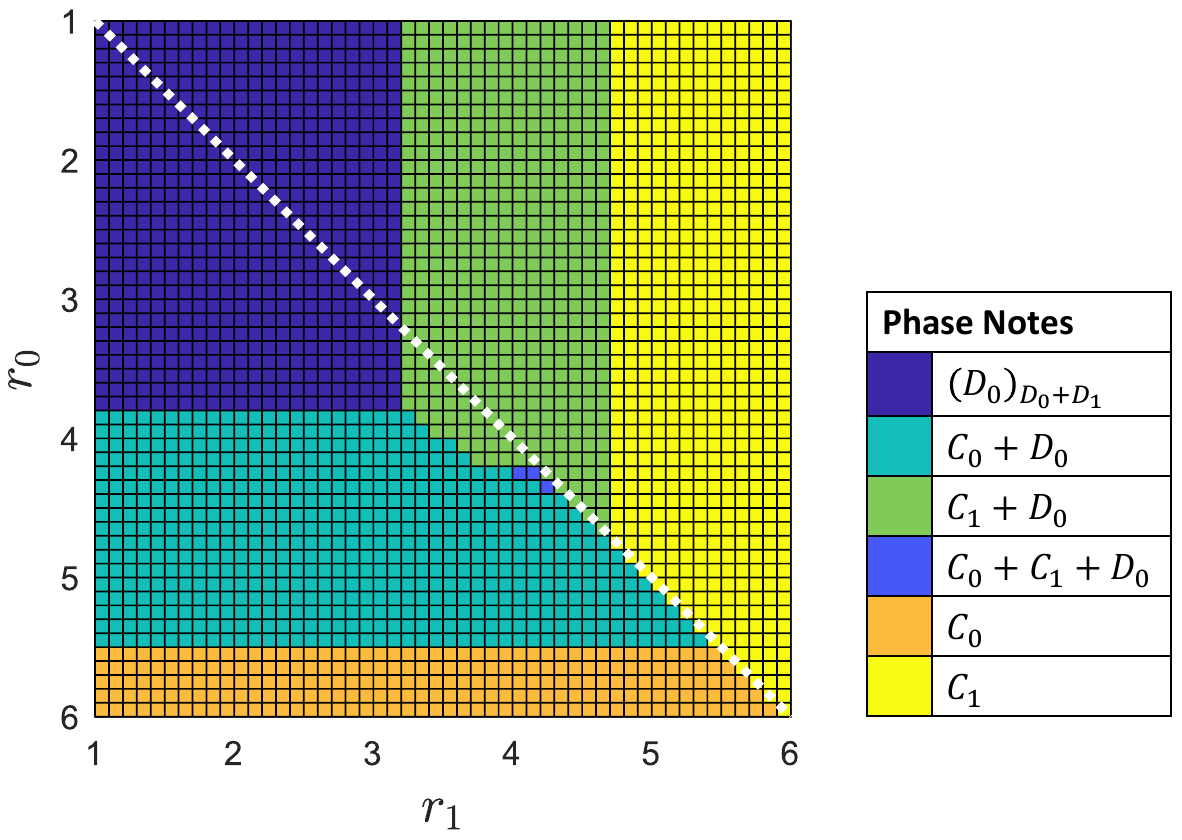}
    \caption{Phase diagram on the $r_0$-$r_1$ parameter plane obtained at other-regarding rate $u=0.5$. The $C_0+C_1+D_0$ phase still exists but is significantly smaller. The white dashed diagonal marks $r_0=r_1$. }\label{fig_phase2D_u05}
\end{figure}

In the above discussed explanation how a three-profile solution emerges, a fundamental point is that self- and other-regarding cooperators are not equally successful in suppressing defectors. In other words, $C_1$ players do it better, and the difference in the portions of $D_0$ players in these domains is the driving force behind the new three-profile solution. This argument can be easily verified if we reduce the effectiveness of other-regarding preferences and explore the phase diagram again. Obviously, if we decrease the other regarding rate to $u=0.5$, the difference between self- and other-regarding preferences becomes smaller. In this case, the other-regarding preference is not completely other-regarding but rather in the middle ground between completely self- and other-regarding. The corresponding phase diagram on the $r_0$-$r_1$ plane is shown in Fig.~\ref{fig_phase2D_u05}. Compared to Fig.~\ref{fig_phase2D_u1}, the changes are clear and they confirm our expectations. First, the critical $r_1$ separating the full defection and two-profile phases shifts towards a higher value, indicating that $C_1$ becomes less strong against defection. In other words, the distribution of self- and other-regarding phases is more symmetrical around the diagonal (white dashed line) in Fig.~\ref{fig_phase2D_u05}. The $C_1+D_0$ phase (light green) still crosses the $r_0 = r_1$ diagonal but not as extensively as for $u=1$, which is also in agreement with our expectation. We also find that the $C_0+C_1+D_0$ solution on the parameter plane is significantly reduced at $u=0.5$. When self- and other-regarding preferences are less distinct, the new solution based on their differences has a smaller chance to emerge.

\subsection{Well-mixed populations}
General evolutionary game dynamics under arbitrary selection noise have been proved unfeasible for simple analytical solutions in structured populations~\cite{ibsen2015computational}, especially under such endogenous behavior-dependent learning rules. However, mathematical results are essential to explore the robustness of our observations. To achieve this, we can consider an infinite and well-mixed population, which is analytically feasible.

\begin{figure}
	\centering
		\includegraphics[width=.95\textwidth]{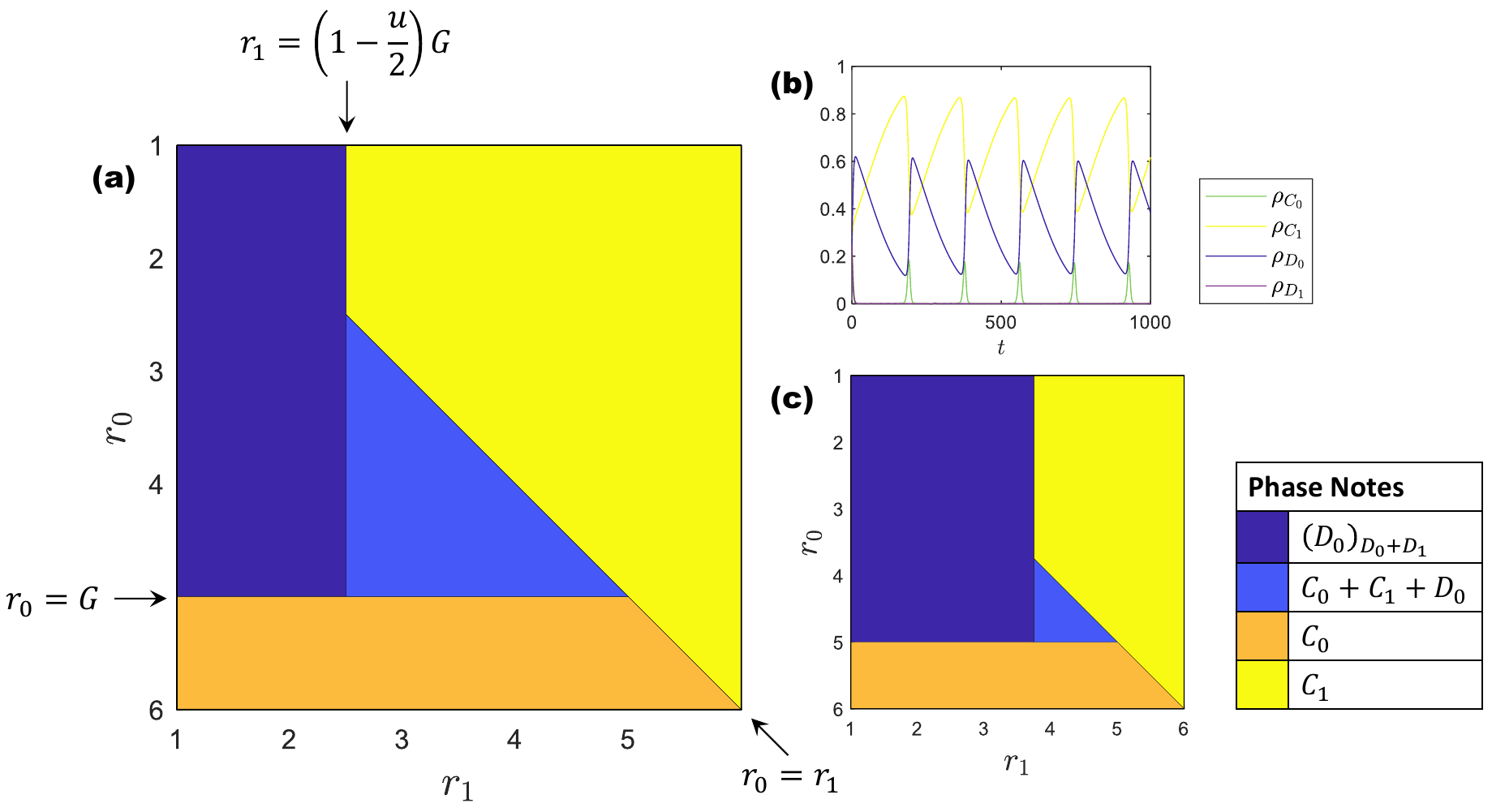}
	\caption{Panels~(a) and (c): phase diagram with respect to self-regarding productivity $r_0$ and other-regarding productivity $r_1$ in an infinite and well-mixed population with any selection noise $\kappa$. Panel~(a): $u=1$. Panel~(c): $u=0.5$. Panel~(b): a typical time evolution pattern of the system state in the $C_0+C_1+D_0$ phase, with parameters $u=1$, $\kappa=0.1$, $r_0=3.5$, $r_1=3.0$. } 
	\label{fig_wm}
\end{figure}

In a well-mixed population, an individual's co-players are randomly selected from the population each time. Individuals interact with these random co-players, playing games and updating profiles. The calculation of self-regarding payoff is the same as in the traditional model. The other-regarding payoff follows our model setting, obtained by averaging the self-regarding payoff that an individual brings to its co-players. Then, different profiles transform into each other at the rate $P_0$ or $\tilde{P}_1$, in the same way as in structured populations. By using the methods of replicator dynamics, we have obtained the corresponding dynamical equations of different profiles' frequencies at a theoretical level (Eq.~(\ref{eqwx_drho}) in \ref{A}). Solving the equilibrium points and analyzing the stability (\ref{A_stability}) lead to theoretical phase diagrams on parameter planes. We have proved that the phase boundaries are consistent across any selection noise $\kappa$. 

The phase diagram on the $r_0$-$r_1$ plane in an infinite and well-mixed population is shown in Fig.~\ref{fig_wm}(a), where we set $u=1$ numerically. The presented phases are derived from the parameter ranges of different stable equilibrium points. In the $(D_0)_{D_0+D_1}$ phase, parameters satisfy $r_0<G$ and $r_1<(1-u/2)G$. The system may equilibrate everywhere where $D_0$ and $D_1$ coexist, but a hypothetical appearance of $C_0$ or $C_1$ leads to the ultimate extinction of $D_1$ and a full $D_0$ state. In the $C_0$ phase, only $C_0$ survives, and the boundaries are $r_0>G$ and $r_0>r_1$. In the $C_1$ phase, only $C_1$ exists, and the phase boundaries are $r_1>(1-u/2)G$ and $r_1>r_0$. In the $C_0+C_1+D_0$ phase, three profiles, $C_0$, $C_1$, and $D_0$ form a dynamic relation, in the parameter range $r_0<G$, $r_1>(1-u/2)G$, and $r_1<r_0$. These analytical phase boundaries are also marked in Fig.~\ref{fig_wm}(a), and we can see that the phase diagram is qualitatively consistent with the previous simulation in structured populations. The effectiveness of $C_1$ players against defection and the emergence of the three-profile solution along the diagonal on the $r_0$-$r_1$ parameter plane are robust and generally valid.

Figure~\ref{fig_wm}(b) shows a typical time evolution of the profile frequencies $\rho_{C_0}$, $\rho_{C_1}$, $\rho_{D_0}$, and $\rho_{D_1}$ in the $C_0+C_1+D_0$ phase. As analyzed in \ref{A_stability}, none of the equilibrium points are stable in this phase, and the system state periodically oscillates around the $C_0+C_1+D_0$ equilibrium. In structured populations, we previously described this phase as dynamic equilibrium, and here we provide a theoretical understanding of it in well-mixed populations. It is also worth noting that the time evolution functions vary with selection noise $\kappa$, but the parameter space over which the periodic oscillation occurs (delineated by phase boundaries) is still independent of $\kappa$.

To complete our study, we show the corresponding phase diagram at $u=0.5$ in Fig.~\ref{fig_wm}(c). We can see that the dynamic $C_0+C_1+D_0$ phase occupies smaller parameter space, similar to the phenomenon revealed in structured populations by Fig.~\ref{fig_phase2D_u05}. Here, the theoretical phase boundary $r_1=(1-u/2)G$ in well-mixed populations switches from $r_1=G/2$ (at $u=1$) to $r_1=3G/4$ (at $u=0.5$), thus shrinking the dominant area of the $C_0+C_1+D_0$ phase.

\section{Conclusion}\label{sec_concl}
Understanding the emergence of cooperation is a fundamental challenge across various scientific disciplines, from social sciences to biology. The most complex explanations involve individuals with cognitive abilities. In such scenarios, solutions may rely on terms and concepts emerging from a long social evolutionary process. Concepts like morals or reputation, which undoubtedly support cooperation, require learning by individuals. The preference to prioritize collective benefits over individual outcomes also represents a sophisticated concept resulting from social learning processes. The primary goal of this study is to demonstrate that such complex behaviors and preferences might evolve spontaneously without the need for additional assumptions.

We utilize a four-profile model, allowing players to choose their worldview. Specifically, when deciding on behavior changes, they can opt for a traditional self-regarding preference, focusing on individual payoff, or an other-regarding preference that aims to optimize the income of others. Unlike approaches that rely on reputation or morals, our model makes no assumptions about the inherent value of these options, avoiding direct support for cooperation or other-regarding preferences. Instead, we let these concepts compete within the diverse parameters of a public goods game. We define the combination of basic strategy and preference as a ``profile'' and track the evolution of these four profiles. Notably, our model does not confine itself to the unrealistic scenario where different preferences operate with identical productivity. By introducing two productivity factors, we explore an expanded model where either self- or other-regarding cooperation is more productive.

Our key findings indicate that other-regarding preferences can emerge and dominate spontaneously, even when self-regarding cooperation is more productive. This suggests that individual and collective interests are not inherently in conflict. An other-regarding player disregards personal income, focusing instead on the welfare of other group members. Yet, in the end, all participants, including the focal individual, achieve higher payoffs than they would under a self-regarding preference. In this way, our simple model is capable of explaining the real-life observations that strongly unselfish acts among competitors could be favored by evolution~\cite{mitteldorf_jtb00,dufwenberg_res11}.

Another intriguing observation is the emergence of a new phase, where self- and other-regarding cooperation coexist with self-regarding defection. This occurs when the productivity of self-regarding cooperation marginally surpasses the other-regarding cooperation, leading to a dynamic equilibrium where both classic solutions engage in ongoing conflict, resulting in stable fractions for three profiles. This dynamic interaction relies on the higher fraction of defection among self-regarding players, allowing other-regarding cooperation to exploit the prevalence of self-regarding defection at the interface. Conversely, due to greater productivity, the self-regarding cooperation profile directly outcompetes other-regarding ones, leading to the reverse process. The robustness of this phenomenon has been confirmed, even in well-mixed populations, through analytical calculations using replicator dynamics.

These results align with research avenues that do not presuppose cooperation-supporting incentives to address the foundational question posed at the start of this paper. Instead, granting individuals the freedom to make choices and approaching the puzzle offers not only a novel approach but also promises broader applicability.

\section*{Acknowledgements}
A.S. was supported by the National Research, Development and Innovation Office (NKFIH) under Grant No. K142948.

\appendix
\section{Well-mixed populations}\label{A}
Here, we supplement the details of theoretical analysis in well-mixed populations. For convenience, we denote the profile set $\mathcal{S}=\{C_0,C_1,D_0,D_1\}$, consisting of all available profiles in the model. We also denote a co-player configuration by $\mathbf{g}=(g_{C_0},g_{C_1},g_{D_0},g_{D_1})$, where $g_{C_0}$, $g_{C_1}$, $g_{D_0}$, and $g_{D_1}$ are the number of one's co-players who employ profile $C_0$, $C_1$, $D_0$, and $D_1$, respectively. In a group of $G$ players, a player has $G-1$ co-players, which means $g_{C_0}+g_{C_1}+g_{D_0}+g_{D_1}=G-1$.

\subsection{Payoff calculation in a single game}
First, we need to express the payoff for each profile in a single game. In a group of $G$ players, containing a player of profile $X\in\mathcal{S}$ and its $G-1$ co-players $\mathbf{g}$, we can write the payoff of the $X$-player based on our model.

\begin{itemize}
    \item Payoff considered by self-regarding preferences
\end{itemize}
According to the description in Sec.~\ref{sec_model}, the self-regarding payoffs $\pi_{X}^{(0)}(\mathbf{g})$ of profile $X\in\mathcal{S}$ under co-player configuration $\mathbf{g}$, which are similar to the ones in the traditional public goods game, are listed as follows:
\begin{subequations}\label{eqwm_pi0}
    \begin{align}
        \pi_{C_0}^{(0)}(\mathbf{g})&=\frac{r_0(g_{C_0}+1)c+r_1 g_{C_1}c}{G}-c, \label{eqwm_pi0C0} \\
        \pi_{C_1}^{(0)}(\mathbf{g})&=\frac{r_0 g_{C_0}c+r_1(g_{C_1}+1)c}{G}-c, \label{eqwm_pi0C1} \\
        \pi_{D_0}^{(0)}(\mathbf{g})=\pi_{D_1}^{(0)}(\mathbf{g})&=\frac{r_0 g_{C_0}c+r_1 g_{C_1}c}{G}. \label{eqwm_pi0D0D1} 
    \end{align}
\end{subequations}

\begin{itemize}
    \item Payoff considered by other-regarding preferences
\end{itemize}
The calculation of other-regarding payoffs is slightly more laborious. According to Sec.~\ref{sec_model}, the other-regarding payoff of a profile $X$ means the average self-regarding payoff that $X$ brings to its co-players. Similar to Eq.~(\ref{eq_orpayoff}) in the main text, we can use the following equation to calculate the other-regarding payoff $\pi_X^{(1)}(\mathbf{g})$ of profile $X$ in a game with co-player configuration $\mathbf{g}$:
\begin{equation}\label{eqwm_orpayoff}
    \pi_X^{(1)}(\mathbf{g})=\frac{\sum_{Y\in\mathcal{S}}g_Y \pi_Y^{(0)}(\mathbf{g}_{-Y,+X})}{G-1}.
\end{equation}
In Eq.~(\ref{eqwm_orpayoff}), $\mathbf{g}_{-Y,+X}$ represents a configuration where $g_Y\gets g_Y-1$, $g_X\gets g_X+1$, which still satisfies $g_{C_0}+g_{C_1}+g_{D_0}+g_{D_1}-1+1=G-1$. For example, $\mathbf{g}_{-C_1,+C_0}=(g_{C_0}+1,g_{C_1}-1,g_{D_0},g_{D_1})$. Intuitively, it means that in $X$'s co-player configuration $\mathbf{g}$, a co-player $Y$'s co-player configuration in the same group can be expressed by modifying $\mathbf{g}$, counting one less $Y$ and one more $X$. Then, $\pi_Y^{(0)}(\mathbf{g}_{-Y,+X})$ is the self-regarding payoff of profile $Y$, $\sum_{Y\in\mathcal{S}}g_Y \pi_Y^{(0)}(\mathbf{g}_{-Y,+X})$ is the total self-regarding payoff of all co-players in $X$'s co-player configuration $\mathbf{g}$, and the average is obtained by dividing by $G-1$.

Applying the general Eq.~(\ref{eqwm_orpayoff}) to $X=C_0$, we can calculate the other-regarding payoff of profile $C_0$ in a single game:
\begin{subequations}\label{eqwm_pi1}
	\begin{align}
		\pi_{C_0}^{(1)}(\mathbf{g})
            =&~\frac{1}{G-1}\left(
            g_{C_0}\pi_{C_0}^{(0)}(\mathbf{g})
            +g_{C_1}\pi_{C_1}^{(0)}(\mathbf{g}_{-C_1,+C_0})
            +g_{D_0}\pi_{D_0}^{(0)}(\mathbf{g}_{-D_0,+C_0})
            +g_{D_1}\pi_{D_1}^{(0)}(\mathbf{g}_{-D_1,+C_0})
            \right)\nonumber\\
            =&~\frac{1}{G-1}\Bigg[
            g_{C_0}\left(\frac{r_0(g_{C_0}+1)c+r_1 g_{C_1}c}{G}-c\right)
            +g_{C_1}\left(\frac{r_0(g_{C_0}+1)c+r_1 g_{C_1}c}{G}-c\right)\nonumber\\
            &+g_{D_0}\frac{r_0(g_{C_0}+1)c+r_1 g_{C_1}c}{G}
            +g_{D_1}\frac{r_0(g_{C_0}+1)c+r_1 g_{C_1}c}{G}
            \Bigg]\nonumber\\
            =&~\frac{r_0(g_{C_0}+1)c+r_1 g_{C_1}c}{G}-\frac{g_{C_0}+g_{C_1}}{G-1}c
		\tag{\ref{eqwm_pi1}{a}} \label{eqwm_pi1C0}.
	\end{align}
\end{subequations}
In Eq.~(\ref{eqwm_pi1C0}), we have utilized the property $\mathbf{g}_{-C_0,+C_0}=(g_{C_0}-1+1,g_{C_1},g_{D_0},g_{D_1})=(g_{C_0},g_{C_1},g_{D_0},g_{D_1})=\mathbf{g}$. Similarly, we can calculate the other-regarding payoff of $C_1$, $D_0$, and $D_1$ in a single game:
\begin{align}
    \pi_{C_1}^{(1)}(\mathbf{g})&=\frac{r_0 g_{C_0}c+r_1(g_{C_1}+1)c}{G}-\frac{g_{C_0}+g_{C_1}}{G-1}c,
    \tag{\ref{eqwm_pi1}{b}} \label{eqwm_pi1C1}
    \\
    \pi_{D_0}^{(1)}(\mathbf{g})=\pi_{D_1}^{(1)}(\mathbf{g})&=\frac{r_0 g_{C_0}c+r_1 g_{C_1}c}{G}-\frac{g_{C_0}+g_{C_1}}{G-1}c.
    \tag{\ref{eqwm_pi1}{c}} \label{eqwm_pi1D0}
\end{align}

\begin{itemize}
    \item Rescaled payoff considered by other-regarding preferences
\end{itemize}
The rescaled other-regarding payoff is the weighting between self- and other-regarding payoffs, as shown in Eq.~(\ref{eq_tildepi1}). Applying Eqs.~(\ref{eqwm_pi0}) and (\ref{eqwm_pi1}), the rescaled other-regarding payoff $\tilde{\pi}_X^{(1)}(\mathbf{g})$ for profile $X\in \mathcal{S}$ in a single game with co-players $\mathbf{g}$ can be written as
\begin{equation}
    \tilde{\pi}_X^{(1)}(\mathbf{g})=(1-u)\pi_X^{(0)}(\mathbf{g})+u\pi_X^{(1)}(\mathbf{g}).
\end{equation}
To specify,
\begin{subequations}\label{eqwx_pitilde1}
    \begin{align}
        \tilde{\pi}_{C_0}^{(1)}(\mathbf{g})&=\frac{r_0(g_{C_0}+1)c+r_1 g_{C_1}c}{G}-\frac{g_{C_0}+g_{C_1}}{G-1}uc-(1-u)c, \label{eqwx_pitilde1C0}
        \\
        \tilde{\pi}_{C_1}^{(1)}(\mathbf{g})&=\frac{r_0 g_{C_0}c+r_1(g_{C_1}+1)c}{G}-\frac{g_{C_0}+g_{C_1}}{G-1}uc-(1-u)c, \label{eqwx_pitilde1C1}
        \\
        \tilde{\pi}_{D_0}^{(1)}(\mathbf{g})=\tilde{\pi}_{D_1}^{(1)}(\mathbf{g})&=\frac{r_0 g_{C_0}c+r_1 g_{C_1}c}{G}-\frac{g_{C_0}+g_{C_1}}{G-1}uc. \label{eqwx_pitilde1D0D1}
    \end{align}
\end{subequations}

\subsection{Statistical mean payoff}
Given the payoff expressions in a single game, we can further calculate the statistical mean payoff of each profile, resulting from multiple games that a player participates each time. In a well-mixed population, the $G-1$ co-players are randomly selected from the population for a focal player. Therefore, we introduce the following function,
\begin{equation}\label{eqwm_f}
    \langle f\rangle=\sum_{\sum_{i\in\mathcal{S}}g_i=G-1}\frac{(G-1)!}{\prod_{i\in\mathcal{S}}g_i!}\left(\prod_{i\in\mathcal{S}}{\rho_i}^{g_i}\right)f(\mathbf{g}),
\end{equation}
which calculates the statistical mean value of function $f(\mathbf{g})$ through all possibilities of configuration $\mathbf{g}$ randomly selected from an infinite well-mixed population. We can apply Eq.~(\ref{eqwm_f}) to calculate the statistical mean self-regarding payoff of profile $C_0$:
\begin{subequations}\label{eqwm_<pi0>}
	\begin{align}
		\langle\pi_{C_0}^{(0)}\rangle
            =&\sum_{\sum_{i\in\mathcal{S}}g_i=G-1}\frac{(G-1)!}{\prod_{i\in\mathcal{S}}g_i!}\left(\prod_{i\in\mathcal{S}}{\rho_i}^{g_i}\right)\pi_{C_0}^{(0)}(\mathbf{g})\nonumber\\
            =&\sum_{g_{C_0}+g_{C_1}+g_{D_0}+g_{D_1}=G-1}\Bigg\{
            \frac{(G-1)!}{g_{C_0}!g_{C_1}!g_{D_0}!g_{D_1}!} 
            {\rho_{C_0}}^{g_{C_0}}{\rho_{C_1}}^{g_{C_1}}{\rho_{D_0}}^{g_{D_0}}{\rho_{D_1}}^{g_{D_1}}\nonumber\\
            &\times
            \left(\frac{r_0(g_{C_0}+1)c+r_1 g_{C_1}c}{G}-c\right)\Bigg\}\nonumber\\
            =&~\frac{r_0 c}{G}(G-1)\rho_{C_0}+\frac{r_1 c}{G}(G-1)\rho_{C_1}+\frac{r_0 c}{G}-c
		\tag{\ref{eqwm_<pi0>}{a}} \label{eqwm_<pi0C0>}.
	\end{align}
\end{subequations}
Similarly, we can apply Eqs.~(\ref{eqwm_pi0C1}) and (\ref{eqwm_pi0D0D1}) to Eq.~(\ref{eqwm_f}) to calculate the statistical mean self-regarding payoff of $C_1$, $D_0$, and $D_1$:
\begin{align}
    \langle\pi_{C_1}^{(0)}\rangle&=\frac{r_0 c}{G}(G-1)\rho_{C_0}+\frac{r_1 c}{G}(G-1)\rho_{C_1}+\frac{r_1 c}{G}-c,
    \tag{\ref{eqwm_<pi0>}{b}} \label{eqwm_<pi0C1>}
    \\
    \langle\pi_{D_0}^{(0)}\rangle=\langle\pi_{D_1}^{(0)}\rangle&=\frac{r_0 c}{G}(G-1)\rho_{C_0}+\frac{r_1 c}{G}(G-1)\rho_{C_1}.
    \tag{\ref{eqwm_<pi0>}{c}} \label{eqwm_<pi0D0>}
\end{align}

We can also use Eq.~(\ref{eqwm_f}) to calculate the statistical mean payoff that a profile brings to its co-players. The ones that we need are the rescaled other-regarding payoffs. Therefore, we apply Eqs.~(\ref{eqwx_pitilde1C0})--(\ref{eqwx_pitilde1D0D1}) to Eq.~(\ref{eqwm_f}) and obtain:
\begin{subequations}
    \begin{align}
        \langle\tilde{\pi}_{C_0}^{(1)}\rangle&=\left(\frac{G-1}{G}r_0-u\right)c\rho_{C_0}+\left(\frac{G-1}{G}r_1-u\right)c\rho_{C_1}+\frac{r_0 c}{G}-(1-u)c, 
        \\
        \langle\tilde{\pi}_{C_1}^{(1)}\rangle&=\left(\frac{G-1}{G}r_0-u\right)c\rho_{C_0}+\left(\frac{G-1}{G}r_1-u\right)c\rho_{C_1}+\frac{r_1 c}{G}-(1-u)c, 
        \\
        \langle\tilde{\pi}_{D_0}^{(1)}\rangle=\langle\tilde{\pi}_{D_1}^{(1)}\rangle&=\left(\frac{G-1}{G}r_0-u\right)c\rho_{C_0}+\left(\frac{G-1}{G}r_1-u\right)c\rho_{C_1}.
    \end{align}
\end{subequations}

\subsection{Profile evolution}
Given the statistical mean payoff calculated, we can further express the evolution of each profile's frequency. The principle of learning rates is the same as modeled in Sec.~\ref{sec_model}. The rates at which profile $X$ learns $Y$ under self- and other-regarding preferences are respectively 
\begin{subequations}
    \begin{align}
        P_0(X\gets Y)&=\frac{1}{1+\exp[-(\langle\pi_Y^{(0)}\rangle-\langle\pi_X^{(0)}\rangle)/\kappa]}, \label{eqwx_P0} \\
        \tilde{P}_1(X\gets Y)&=\frac{1}{1+\exp[-(\langle\tilde{\pi}_Y^{(1)}\rangle-\langle\tilde{\pi}_X^{(1)}\rangle)/\kappa]}. \label{eqwx_P1}
    \end{align}
\end{subequations}
If profile $X$ is self-regarding (i.e., $X\in\{C_0,D_0\}$), then Eq.~(\ref{eqwx_P0}) is used. If profile $X$ is other-regarding (i.e., $X\in\{C_1,D_1\}$), then Eq.~(\ref{eqwx_P1}) is used.

For simplicity, we denote the following function:
\begin{equation}
    F(x)=\frac{1}{1+\exp(-x/\kappa)},
\end{equation}
which has two properties: (1) $F(x_1)>F(x_2)\Leftrightarrow x_1>x_2$; (2) $F(-x)=1-F(x)$.

The replicator equation of the system can then be expressed as follows:
\begin{subequations}\label{eqwx_drho}
    \begin{align}
        \dot{\rho}_{C_0}=&~
        \rho_{C_1}\rho_{C_0}\tilde{P}_1(C_1\gets C_0)
        +\rho_{D_0}\rho_{C_0}P_0(D_0\gets C_0)
        +\rho_{D_1}\rho_{C_0}\tilde{P}_1(D_1\gets C_0)
        \nonumber\\
        &-\rho_{C_0}\rho_{C_1}P_0(C_0\gets C_1)
        -\rho_{C_0}\rho_{D_0}P_0(C_0\gets D_0)
        -\rho_{C_0}\rho_{D_1}P_0(C_0\gets D_1)
        \nonumber\\
        =&~\rho_{C_0}\rho_{C_1}\left[F\left(\frac{c}{G}(r_0-r_1)\right)-F\left(-\frac{c}{G}(r_0-r_1)\right)\right]
        \nonumber\\
        &+\rho_{C_0}\rho_{D_0}\left[F\left(\frac{c}{G}r_0-c\right)-F\left(-\frac{c}{G}r_0+c\right)\right]
        \nonumber\\
        &+\rho_{C_0}(1-\rho_{C_0}-\rho_{C_1}-\rho_{D_0})\left[F\left(\frac{c}{G}r_0-(1-u)c\right)-F\left(-\frac{c}{G}r_0+c\right)\right], \label{eqwx_drhoC0}
        \\
        \dot{\rho}_{C_1}=&~
        \rho_{C_0}\rho_{C_1}P_0(C_0\gets C_1)
        +\rho_{D_0}\rho_{C_1}P_0(D_0\gets C_1)
        +\rho_{D_1}\rho_{C_1}\tilde{P}_1(D_1\gets C_1)
        \nonumber\\
        &-\rho_{C_1}\rho_{C_0}\tilde{P}_1(C_1\gets C_0)
        -\rho_{C_1}\rho_{D_0}\tilde{P}_1(C_1\gets D_0)
        -\rho_{C_1}\rho_{D_1}\tilde{P}_1(C_1\gets D_1)
        \nonumber\\
        =&~\rho_{C_1}\rho_{C_0}\left[F\left(-\frac{c}{G}(r_0-r_1)\right)-F\left(\frac{c}{G}(r_0-r_1)\right)\right]
        \nonumber\\
        &+\rho_{C_1}\rho_{D_0}\left[F\left(\frac{c}{G}r_1-c\right)-F\left(-\frac{c}{G}r_1+(1-u)c\right)\right]
        \nonumber\\
        &+\rho_{C_1}(1-\rho_{C_0}-\rho_{C_1}-\rho_{D_0})\left[F\left(\frac{c}{G}r_1-(1-u)c\right)-F\left(-\frac{c}{G}r_1+(1-u)c\right)\right], \label{eqwx_drhoC1}
        \\
        \dot{\rho}_{D_0}=&~
        \rho_{C_0}\rho_{D_0}P_0(C_0\gets D_0)
        +\rho_{C_1}\rho_{D_0}\tilde{P}_1(C_1\gets D_0)
        +\rho_{D_1}\rho_{D_0}\tilde{P}_1(D_1\gets D_0)
        \nonumber\\
        &-\rho_{D_0}\rho_{C_0}P_0(D_0\gets C_0)
        -\rho_{D_0}\rho_{C_1}P_0(D_0\gets C_1)
        -\rho_{D_0}\rho_{D_1}P_0(D_0\gets D_1)
        \nonumber\\
        =&~\rho_{D_0}\rho_{C_0}\left[F\left(-\frac{c}{G}r_0+c)\right)-F\left(\frac{c}{G}r_0-c)\right)\right]
        \nonumber\\
        &+\rho_{D_0}\rho_{C_1}\left[F\left(-\frac{c}{G}r_1+(1-u)c\right)-F\left(\frac{c}{G}r_1-c\right)\right].
    \end{align}
\end{subequations}
A profile is adopted with certain rates when two players meet, which causes the evolution of $\rho_{C_0}$, $\rho_{C_1}$, and $\rho_{D_0}$. For example, $\rho_{C_1}\rho_{C_0}\tilde{P}_1(C_1\gets C_0)$ in $\dot{\rho}_{C_0}$ means the transformation from $C_1$ to $C_0$ with other-regarding learning rate $\tilde{P}_1(C_1\gets C_0)$, for which $\rho_{C_0}$ increases. Similarly, $-\rho_{C_0}\rho_{C_1}P_0(C_0\gets C_1)$ in $\dot{\rho}_{C_0}$ represents the transformation from $C_0$ to $C_1$ with self-regarding learning rate $P_0(C_0\gets C_1)$, for which $\rho_{C_0}$ decreases. 

Note that the degrees of freedom in system~(\ref{eqwx_drho}) are $3$, because of the constraint $\rho_{C_0}+\rho_{C_1}+\rho_{D_0}+\rho_{D_1}=1$ and its consequence $\dot{\rho}_{D_1}=-\dot{\rho}_{C_0}-\dot{\rho}_{C_1}-\dot{\rho}_{D_0}$. One can easily verify that $\dot{\rho}_{D_1}\leq 0$ always holds, thus $D_1$ cannot survive in evolution.

\subsection{Stability analysis}\label{A_stability}
Denote the system state by $\boldsymbol{\rho}=(\rho_{C_0},\rho_{C_1},\rho_{D_0},\rho_{D_1})$. Solving $\dot{\rho}_{C_0}=0$, $\dot{\rho}_{C_1}=0$, $\dot{\rho}_{D_0}=0$ in Eq.~(\ref{eqwx_drho}) yields the equilibrium points, denoted by $\boldsymbol{\rho}^*$. For simplicity, we use the same notation $\boldsymbol{\rho}^*$ to refer to each equilibrium point, with the understanding that each notation is contextually defined and marked by $\bullet$. To study stability of the equilibrium points, we can compute the Jacobian matrix:
\begin{equation}\label{eqwx_J}
    J=
    \begin{pmatrix}
        \displaystyle{\frac{\partial\dot{\rho}_{C_0}}{\partial \rho_{C_0}}} & 
        \displaystyle{\frac{\partial\dot{\rho}_{C_0}}{\partial \rho_{C_1}}} & 
        \displaystyle{\frac{\partial\dot{\rho}_{C_0}}{\partial \rho_{D_0}}} \\[1em]
        \displaystyle{\frac{\partial\dot{\rho}_{C_1}}{\partial \rho_{C_0}}} & 
        \displaystyle{\frac{\partial\dot{\rho}_{C_1}}{\partial \rho_{C_1}}} & 
        \displaystyle{\frac{\partial\dot{\rho}_{C_1}}{\partial \rho_{D_0}}} \\[1em]
        \displaystyle{\frac{\partial\dot{\rho}_{D_0}}{\partial \rho_{C_0}}} & 
        \displaystyle{\frac{\partial\dot{\rho}_{D_0}}{\partial \rho_{C_1}}} & 
        \displaystyle{\frac{\partial\dot{\rho}_{D_0}}{\partial \rho_{D_0}}} 
    \end{pmatrix}.
\end{equation}
If the Jacobian is negative definite, then an equilibrium point is stable; otherwise, it is unstable.

Below, we list and discuss each equilibrium after solving $\dot{\rho}_{C_0}=0$, $\dot{\rho}_{C_1}=0$, $\dot{\rho}_{D_0}=0$. For simplicity, after checking their existence, we only show the existing ones here.

\begin{itemize}
    \item The $C_0$ equilibrium: $\boldsymbol{\rho}^*=(1,0,0,0)$.
\end{itemize}
One solution is $\rho_{C_0}=1$, where profile $C_0$ dominates. Substituting $\boldsymbol{\rho}^*=(1,0,0,0)$ into Eq.~(\ref{eqwx_J}), we have the Jacobian matrix at this equilibrium point:
\begin{equation}
    J|_{\boldsymbol{\rho}=\boldsymbol{\rho}^*}=
    \begin{pmatrix}
        J_{11} & J_{12} & J_{13} \\
        0 & J_{22} & 0 \\
        0 & 0 & J_{33}
    \end{pmatrix},
\end{equation}
where
\begin{subequations}
    \begin{align}
        J_{11}&=-F\left(\frac{c}{G}r_0-(1-u)c\right)+F\left(-\frac{c}{G}r_0+c\right), \\
        J_{22}&=F\left(-\frac{c}{G}(r_0-r_1)\right)-F\left(\frac{c}{G}(r_0-r_1)\right), \\
        J_{33}&=F\left(-\frac{c}{G}r_0+c\right)-F\left(\frac{c}{G}r_0-c\right).
    \end{align}
\end{subequations}

The conditions of $J|_{\boldsymbol{\rho}=\boldsymbol{\rho}^*}$ being negative definite are
\begin{subequations}
    \begin{align}
        J_{11}<0 \Leftrightarrow
        -\frac{c}{G}r_0+c<\frac{c}{G}r_0-(1-u)c &\Leftrightarrow
        r_0>\left(1-\frac{u}{2}\right)G, \label{eqwx_C0J11}
        \\
        J_{22}<0 \Leftrightarrow
        -\frac{c}{G}(r_0-r_1)<\frac{c}{G}(r_0-r_1) &\Leftrightarrow
        r_0>r_1,
        \\
        J_{33}<0 \Leftrightarrow
        -\frac{c}{G}r_0+c<\frac{c}{G}r_0-c &\Leftrightarrow
        r_0>G. \label{eqwx_C0J33}
    \end{align}
\end{subequations}
Since $0\leq u\leq 1$, the constraint in Eq.~(\ref{eqwx_C0J33}) is stronger than in Eq.~(\ref{eqwx_C0J11}). To summarize, the conditions of the $C_0$ equilibrium being stable are $r_0>G$ and $r_0>r_1$.

\begin{itemize}
    \item The $C_1$ equilibrium: $\boldsymbol{\rho}^*=(0,1,0,0)$.
\end{itemize}
Another solution is $\rho_{C_1}=1$, where profile $C_1$ dominates. Substituting $\boldsymbol{\rho}^*=(0,1,0,0)$ into Eq.~(\ref{eqwx_J}), we have this equilibrium point's Jacobian matrix:
\begin{equation}
    J|_{\boldsymbol{\rho}=\boldsymbol{\rho}^*}=
    \begin{pmatrix}
        J_{11} & 0 & 0 \\
        J_{21} & J_{22} & J_{23} \\
        0 & 0 & J_{33}
    \end{pmatrix},
\end{equation}
where
\begin{subequations}
    \begin{align}
        J_{11}&=F\left(\frac{c}{G}(r_0-r_1)\right)-F\left(-\frac{c}{G}(r_0-r_1)\right), \\
        J_{22}&=-F\left(\frac{c}{G}r_1-(1-u)c\right)+F\left(-\frac{c}{G}r_1+(1-u)c\right), \\
        J_{33}&=F\left(-\frac{c}{G}r_1+(1-u)c\right)-F\left(\frac{c}{G}r_1-c\right).
    \end{align}
\end{subequations}

The conditions of $J|_{\boldsymbol{\rho}=\boldsymbol{\rho}^*}$ being negative definite are
\begin{subequations}
    \begin{align}
        J_{11}<0 \Leftrightarrow
        \frac{c}{G}(r_0-r_1)<-\frac{c}{G}(r_0-r_1) &\Leftrightarrow
        r_1>r_0,
        \\
        J_{22}<0 \Leftrightarrow
        -\frac{c}{G}r_1+(1-u)c<\frac{c}{G}r_1-(1-u)c &\Leftrightarrow
        r_1>(1-u)G, \label{eqwx_C1J22}
        \\
        J_{33}<0 \Leftrightarrow
        -\frac{c}{G}r_1+(1-u)c<\frac{c}{G}r_1-c &\Leftrightarrow
        r_1>\left(1-\frac{u}{2}\right)G. \label{eqwx_C1J33}
    \end{align}
\end{subequations}
We can see that the constraint in Eq.~(\ref{eqwx_C1J33}) is stronger than in Eq.~(\ref{eqwx_C1J22}). To summarize, the conditions of the $C_1$ equilibrium being stable are $r_1>(1-u/2)G$ and $r_1>r_0$.

\begin{itemize}
    \item The $D_0+D_1$ equilibrium line: $\boldsymbol{\rho}^*=(0,0,\rho_{D_0}^*,1-\rho_{D_0}^*)$, where $0\leq \rho_{D_0}^*\leq 1$.
\end{itemize}
This solution is not a point, but a line. $D_0$ and $D_1$ can coexist and equilibrate everywhere on this equilibrium line. To study this type of equilibrium, we can treat the two variables on the equilibrium line as a whole, as some previous work did~\cite{wang2024evolution}. In this equilibrium, we treat $D_0$ and $D_1$ as whole. That is, there are three variables in the system: $C_0$, $C_1$, and $D_0+D_1$. Furthermore, due to the constraint $C_0+C_1+(D_0+D_1)=1$, the degrees of freedom decrease to 2. 

Therefore, the system can be described by $\dot{\rho}_{C_0}$ and $\dot{\rho}_{C_1}$ only, as given by Eqs.~(\ref{eqwx_drhoC0}) and (\ref{eqwx_drhoC1}). The Jacobian matrix of this system at this equilibrium is
\begin{equation}
    J|_{\boldsymbol{\rho}=\boldsymbol{\rho}^*}=
    \left.\begin{pmatrix}
        \displaystyle{\frac{\partial\dot{\rho}_{C_0}}{\partial\rho_{C_0}}} & \displaystyle{\frac{\partial\dot{\rho}_{C_0}}{\partial\rho_{C_1}}} \\[1em]
        \displaystyle{\frac{\partial\dot{\rho}_{C_1}}{\partial\rho_{C_0}}} &
        \displaystyle{\frac{\partial\dot{\rho}_{C_1}}{\partial\rho_{C_1}}}
    \end{pmatrix}
    \right|_{\boldsymbol{\rho}=\boldsymbol{\rho}^*}=
    \begin{pmatrix}
        J_{11} & 0 \\
        0 & J_{22}
    \end{pmatrix},
\end{equation}
where
\begin{subequations}\label{eqwx_JD0D1}
    \begin{align}
        J_{11}=&~\rho_{D_0}^*\left[F\left(\frac{c}{G}r_0-c\right)-F\left(\frac{c}{G}r_0-(1-u)c\right)\right] \nonumber\\
        &+F\left(\frac{c}{G}r_0-(1-u)c\right)-F\left(-\frac{c}{G}r_0+c\right) \nonumber\\
        \geq&~F\left(\frac{c}{G}r_0-c\right)-F\left(-\frac{c}{G}r_0+c\right), \\
        J_{22}=&~\rho_{D_0}^*\left[F\left(\frac{c}{G}r_1-c\right)-F\left(\frac{c}{G}r_1-(1-u)c\right)\right] \nonumber\\
        &+F\left(\frac{c}{G}r_1-(1-u)c\right)-F\left(-\frac{c}{G}r_1+(1-u)c\right) \nonumber\\
        \geq&~F\left(\frac{c}{G}r_1-c\right)-F\left(-\frac{c}{G}r_1+(1-u)c\right).
    \end{align}
\end{subequations}
The inequality deflation in Eq.~(\ref{eqwx_JD0D1}) used $\rho_{D_0}^*\leq 1$.

If the system is stable at any point on the $D_0+D_1$ equilibrium line, then we say that this equilibrium is stable. In this way, we check the conditions that there exists $\rho_{D_0}^*$ to let $J|_{\boldsymbol{\rho}=\boldsymbol{\rho}^*}$ be negative definite. Applying Eq.~(\ref{eqwx_JD0D1}), we have 
\begin{equation}
    \left\{
    \begin{array}{@{\hspace{0.1em}}l@{\hspace{0.25em}}l}
    J_{11} &<0 \\[0.1em]
    J_{22} &<0 
    \end{array}
    \right.
    ,~\exists \rho_{D_0}^* \Leftrightarrow
    \left\{
    \begin{array}{@{\hspace{0.1em}}l@{\hspace{0.25em}}l}
    \displaystyle{\frac{c}{G}r_0-c}&<\displaystyle{-\frac{c}{G}r_0+c} \\[1em]
    \displaystyle{\frac{c}{G}r_1-c}&<\displaystyle{-\frac{c}{G}r_1+(1-u)c}
    \end{array}
    \right.
    \Leftrightarrow
    \left\{
    \begin{array}{@{\hspace{0.1em}}l@{\hspace{0.25em}}l}
    r_0 &<G, \\[0.5em]
    r_1 &<\displaystyle{\left(1-\frac{u}{2}\right)G}.
    \end{array}
    \right.
\end{equation}
To summarize, the conditions of the $D_0+D_1$ equilibrium being stable are $r_0<G$ and $r_1<(1-u/2)G$.

\begin{itemize}
    \item The $C_0+C_1+D_0$ equilibrium: $\boldsymbol{\rho}^*=(\rho_{C_0}^*,\rho_{C_1}^*,\rho_{D_0}^*,0)$.
\end{itemize}
In this solution, $\rho_{C_0}^*,\rho_{C_1}^*,\rho_{D_0}^*\neq 0$, which can be obtained by solving 
\begin{subequations}
    \begin{align}
        0=&~\rho_{C_1}\left[F\left(\frac{c}{G}(r_0-r_1)\right)-F\left(-\frac{c}{G}(r_0-r_1)\right)\right]+\rho_{D_0}\left[F\left(\frac{c}{G}r_0-c\right)-F\left(-\frac{c}{G}r_0+c\right)\right]
        \nonumber\\
        &+(1-\rho_{C_0}-\rho_{C_1}-\rho_{D_0})\left[F\left(\frac{c}{G}r_0-(1-u)c\right)-F\left(-\frac{c}{G}r_0+c\right)\right],
        \\
        0=&~\rho_{C_0}\left[F\left(-\frac{c}{G}(r_0-r_1)\right)-F\left(\frac{c}{G}(r_0-r_1)\right)\right]+\rho_{D_0}\left[F\left(\frac{c}{G}r_1-c\right)-F\left(-\frac{c}{G}r_1+(1-u)c\right)\right]
        \nonumber\\
        &+(1-\rho_{C_0}-\rho_{C_1}-\rho_{D_0})\left[F\left(\frac{c}{G}r_1-(1-u)c\right)-F\left(-\frac{c}{G}r_1+(1-u)c\right)\right],
        \\
        0=&~\rho_{C_0}\left[F\left(-\frac{c}{G}r_0+c)\right)-F\left(\frac{c}{G}r_0-c)\right)\right]+\rho_{C_1}\left[F\left(-\frac{c}{G}r_1+(1-u)c\right)-F\left(\frac{c}{G}r_1-c\right)\right]. \label{eqwx_drhoD0=0}
    \end{align}
\end{subequations}
According to Eq.~(\ref{eqwx_drhoD0=0}), 
\begin{equation}
    \frac{\partial\dot{\rho}_{D_0}}{\partial\rho_{D_0}}=\rho_{C_0}\left[F\left(-\frac{c}{G}r_0+c)\right)-F\left(\frac{c}{G}r_0-c)\right)\right]+\rho_{C_1}\left[F\left(-\frac{c}{G}r_1+(1-u)c\right)-F\left(\frac{c}{G}r_1-c\right)\right]=0,
\end{equation}
which means that the Jacobian matrix at this equilibrium is not negative definite. Therefore, the $C_0+C_1+D_0$ equilibrium point is not stable.

\section*{References}
%  \bibliographystyle{iop}
% \bibliography{ref.bib}

\begin{thebibliography}{10}
\expandafter\ifx\csname url\endcsname\relax
  \def\url#1{{\tt #1}}\fi
\expandafter\ifx\csname urlprefix\endcsname\relax\def\urlprefix{URL }\fi
\providecommand{\eprint}[2][]{\url{#2}}
% Bibliography created with iopart-num v2.1
% /biblio/bibtex/contrib/iopart-num

\bibitem{weibull_95}
Weibull J~W 1995 {\em Evolutionary game theory\/} (Cambridge, MA: MIT Press)

\bibitem{sigmund2010calculus}
Sigmund K 2010 {\em The calculus of selfishness\/} (Princeton University Press)

\bibitem{perc_jrsi13}
Perc M, G{\'o}mez-Garde{\~n}es J, Szolnoki A and Flor{\'{\i}a and Y Moreno} L~M
  2013 {\em J. R. Soc. Interface\/} {\bf 10} 20120997

\bibitem{javarone_epl16}
Javarone M~A, Antonioni A and Caravelli F 2016 {\em EPL\/} {\bf 114} 38001

\bibitem{wang2022reversed}
Wang C and Szolnoki A 2022 {\em New J. Phys.\/} {\bf 24} 123030

\bibitem{maynard_82}
Maynard~Smith J 1982 {\em Evolution and the theory of games\/} (Cambridge,
  U.K.: Cambridge University Press)

\bibitem{hardin1968tragedy}
Hardin G 1968 {\em Science\/} {\bf 162} 1243--1248

\bibitem{nowak2006evolutionary}
Nowak M~A 2006 {\em Evolutionary dynamics: exploring the equations of life\/}
  (Harvard University Press)

\bibitem{nowak_11}
Nowak M~A and Highfield R 2011 {\em Supercooperators: Altruism, evolution, and
  why we need each other to succeed\/} (New York: Free Press)

\bibitem{nowak_n93}
Nowak M~A and Sigmund K 1993 {\em Nature\/} {\bf 364} 56--58

\bibitem{press_pnas12}
Press W and Dyson F 2012 {\em Proc. Natl. Acad. Sci.\/} {\bf 109}
  10409--10413

\bibitem{nowak2006five}
Nowak M~A 2006 {\em Science\/} {\bf 314} 1560--1563

\bibitem{amaral_pre18}
Amaral M~A and Javarone M~A 2018 {\em Phys. Rev. E\/} {\bf 97} 042305

\bibitem{cardinot_njp19}
Cardinot M, O'Riordan C, Griffith J and Szolnoki A 2019 {\em New J. Phys.\/}
  {\bf 21} 073038

\bibitem{chen_xj_pa08}
Chen X, Fu F and Wang L 2008 {\em Physica A\/} {\bf 387} 5609--5615

\bibitem{wang_cq_amc23}
Wang C and Szolnoki A 2023 {\em Appl. Math. Comput.\/} {\bf 449} 127941

\bibitem{liang_rh_pre22}
Liang R, Wang Q, Zhang J, Zheng G, Ma L and Chen L 2022 {\em Phys. Rev. E\/}
  {\bf 105} 054302

\bibitem{helbing_ploscb10}
Helbing D, Szolnoki A, Perc M and Szab{\'o} G 2010 {\em PLoS Comput. Biol.\/}
  {\bf 6} e1000758

\bibitem{brandt_pnas06}
Brandt H, Hauert C and Sigmund K 2006 {\em Proc. Natl. Acad. Sci.\/} {\bf
  103} 495--497

\bibitem{javarone_jsm24}
Javarone M~A and Singh S~P 2024 {\em J. Stat. Mech.\/} {\bf 2024} 023404

\bibitem{szolnoki_epl10}
Szolnoki A and Perc M 2010 {\em EPL\/} {\bf 92} 38003

\bibitem{nowak1992evolutionary}
Nowak M~A and May R~M 1992 {\em Nature\/} {\bf 359} 826--829

\bibitem{santos_prl05}
Santos F~C and Pacheco J~M 2005 {\em Phys. Rev. Lett.\/} {\bf 95} 098104

\bibitem{fu_pla07}
Fu F, Chen X, Liu L and Wang L 2007 {\em Phys. Lett. A\/} {\bf 371} 58--64

\bibitem{floria_pre09}
Flor{\'{\i}}a L~M, Gracia-L{\'a}zaro C, G{\'o}mez-Garde{\~n}es J and Moreno Y
  2009 {\em Phys. Rev. E\/} {\bf 79} 026106

\bibitem{wang_z_epjb15}
Wang Z, Wang L, Szolnoki A and Perc M 2015 {\em Eur. Phys. J. B\/} {\bf 88} 124

\bibitem{lieberman2005evolutionary}
Lieberman E, Hauert C and Nowak M~A 2005 {\em Nature\/} {\bf 433} 312--316

\bibitem{allen2017evolutionary}
Allen B, Lippner G, Chen Y~T, Fotouhi B, Momeni N, Yau S~T and Nowak M~A 2017
  {\em Nature\/} {\bf 544} 227--230

\bibitem{szabo_pr07}
Szab{\'o} G and F{\'a}th G 2007 {\em Phys. Rep.\/} {\bf 446} 97--216

\bibitem{perc2017statistical}
Perc M, Jordan J~J, Rand D~G, Wang Z, Boccaletti S and Szolnoki A 2017 {\em
  Phys. Rep.\/} {\bf 687} 1--51

\bibitem{frohlich_pc04}
Frohlich N, Oppenheimer J and Kurki A 2004 {\em Publ. Choice\/} {\bf 119}
  91--117

\bibitem{szabo_jtb12}
Szab{\'o} G and Szolnoki A 2012 {\em J. Theor. Biol.\/} {\bf 299} 81--87

\bibitem{pei_hy_cpb21}
Pei H, Yan G and Wang H 2021 {\em Chin. Phys. B\/} {\bf 30} 050203

\bibitem{burkart_pnas07}
Burkart J~M, Fehr E, Efferson C and van Schaik C~P 2007 {\em Proc. Natl. Acad.
  Sci.\/} {\bf 104} 19762--19766

\bibitem{grund_srep13}
Grund T, Waloszek C and Helbing D 2013 {\em Sci. Rep.\/} {\bf 3} 1480

\bibitem{platkowski_pa22}
Platkowski T 2022 {\em Physica A\/} {\bf 599} 127403

\bibitem{han2023novel}
Han Z, Zhu P and Shi J 2023 {\em Chaos, Solit. and Fract.\/} {\bf 166} 112894

\bibitem{mitteldorf_jtb00}
Mitteldorf J and Wilson D~S 2000 {\em J. Theor. Biol.\/} {\bf 204} 481--496

\bibitem{dufwenberg_res11}
Dufwenberg M, Heidhues P, Kirchsteiger G, Riedel F and Sobel J 2011 {\em Rev.
  Econ. Stud.\/} {\bf 78} 613--639

\bibitem{wang_z_pa11}
Wang Z, Du W~B, Cao X~B and Zhang L~Z 2011 {\em Physica A\/} {\bf 390}
  1234--1239

\bibitem{li_q_c22}
Li Q, Li S, Zhang Y, Chen X and Yang S 2022 {\em Chaos\/} {\bf 32} 113117

\bibitem{zhang_yl_tcss24}
Zhang Y, Li Y, Chen X and Xie G 2024 {\em IEEE Trans. Comput. Soc. Syst.\/}  in
  press

\bibitem{szolnoki_pre09c}
Szolnoki A, Perc M and Szab{\'o} G 2009 {\em Phys. Rev. E\/} {\bf 80} 056109

\bibitem{flores_jtb21}
Flores L~S, Fernandes H~C, Amaral M~A and Vainstein M~H 2021 {\em J. Theor.
  Biol.\/} {\bf 524} 110737

\bibitem{quan_j_csf21}
Quan J, Pu Z and Wang X 2021 {\em Chaos, Solit. and Fract.\/} {\bf 151} 111229

\bibitem{wang_jw_pla22}
Wang J, Dai W, He J, Yu F and Shen X 2022 {\em Phys. Lett. A\/} {\bf 447}
  128302

\bibitem{yu_fy_csf22}
Yu F, Wang J and He J 2022 {\em Chaos, Solit. and Fract.\/} {\bf 165} 112755

\bibitem{sun_xp_pla23}
Sun X, Han L, Wang M, Liu S and Shen Y 2023 {\em Phys. Lett. A\/} {\bf 474}
  128837

\bibitem{szabo1998evolutionary}
Szab{\'o} G and T{\H{o}}ke C 1998 {\em Phys. Rev. E\/} {\bf 58} 69

\bibitem{flores_pre23}
Flores L~S, Vainstein M~H, Fernandes H~C~M and Amaral M~A 2023 {\em Phys. Rev.
  E\/} {\bf 108} 024111

\bibitem{zhang_h_csf23}
Zhang H 2023 {\em Chaos, Solit. and Fract.\/} {\bf 174} 113874

\bibitem{duan_yx_csf23}
Duan Y, Huang J and Zhang J 2023 {\em Chaos, Solit. and Fract.\/} {\bf 174}
  113862

\bibitem{quan_j_c23}
Quan J, Zhang X, Chen W and Wang X 2023 {\em Chaos\/} {\bf 33} 073107

\bibitem{roca_prl06}
Roca C~P, Cuesta J~A and S{\'a}nchez A 2006 {\em Phys. Rev. Lett.\/} {\bf 97}
  158701

\bibitem{szolnoki_epjb09}
Szolnoki A and Perc M 2009 {\em Eur. Phys. J. B\/} {\bf 67} 337--344

\bibitem{ibsen2015computational}
Ibsen-Jensen R, Chatterjee K and Nowak M~A 2015 {\em Proc. Natl. Acad. Sci.\/} {\bf 112} 15636--15641

\bibitem{wang2024evolution}
Wang C 2024 {\em Appl. Math. Comput.\/} {\bf 471} 128595

\end{thebibliography}

\providecommand{\newblock}{}

\end{document}